\documentclass[reprint,
 amsmath,amssymb,
aps,
prd,
showpacs
]{revtex4-1}
\usepackage{amsmath}
\usepackage{physics}
\usepackage{amssymb}
\usepackage{amsbsy}
\usepackage{xcolor}

\usepackage{graphicx}
\usepackage{dcolumn}
\usepackage{bm}

\begin{document}

\preprint{APS/123-QED}

\date{\today}
\title{ de Sitter 
Fractional Quantum Cosmology}
\author{S. Jalalzadeh$^{a1}$, E.W. Oliveira Costa$^{a2}$ and P.V. Moniz$^{b3}$}
\affiliation{$^{a}$ {\it Departamento de F\'{i}sica, Universidade Federal de Pernambuco,
Recife, PE, 52171-900, Brazil}\\
$^b$ {\it Departmento de F\'{i}sica, }
Centro de Matem\'atica e Aplica\c c\~oes (CMA-UBI),\\ Universidade da Beira Interior, 6200 Covilh\~a, Portugal\\
$^1${\tt shahram.jalalzadeh@ufpe.br}\\
$^2${\tt emanuel.wcosta@ufpe.br}\\
$^3${\tt pmoniz@ubi.pt}\\}

\date{\today}
\vskip -.5truecm
\begin{abstract}
We employ Riesz's fractional derivative into the Wheeler--DeWitt equation for a closed de Sitter geometry and obtain the no-boundary and tunneling wavefunctions. From the corresponding probability distributions,  the event horizon of the nucleated universe can be a fractal surface with dimensions between $2\leq D<3$. Concretely, the tunneling wavefunction favors fractal dimensions less than $2.5$ and an accelerated power-law phase. Differently, the no-boundary proposal conveys  fractal  dimensions close to $3$, with  the universe  instead entering a decelerated phase. 
Subsequently, we extend our discussion towards (non-trivial compact) flat and open scenarios. Results suggest that given the probability of creation of a closed inflationary universe in the tunneling proposal is exponentially suppressed, a flat or an open universe becomes favored within  fractional inflationary quantum universe.

\end{abstract}
\maketitle




\section{Introduction}

The entire visible universe appears to be highly unique in a way that is not suggested solely by known dynamical laws.  
Being more specific, from displaying a small large-scale curvature and approximate homogeneity and isotropy of the matter distribution on the largest current scales, the universe seems to demand judicious boundary conditions for its emergence. Namely, an exceptionally high degree of order in its very early stage. From such, it will then be evolving with increasing entropy, aligned with the second law of thermodynamics. The Hartle--Hawking `no-boundary' proposal \cite{Hawking1,Hawking12} and the {Linde--Vilenkin `tunneling' proposal \cite{1984ZhETF,Linde:1983mx,Linde1984,Vilenkin1,Vilenkin14,Vilenkin15,Vilenkin16}} are two popular proposals for quantum states of the model universes that accommodate this perspective \footnote{Let us remind that Vilenkin's original 1982 paper \cite{Vilenkin1} claimed that $P \sim e^{S}$, thus Vilenkin's original result coincided with Hartle and Hawking's subsequent 1983 result \cite{Hawking1}. Linde first obtained the result $P \sim e^{-S}$ in a series of papers, beginning with his first article \cite{1984ZhETF} and continuing with his subsequent papers \cite{Linde:1983mx,Linde1984}. It was stated there that Vilenkin's original result was incorrect. Vilenkin came to the same conclusion a few months later. Thus, the $P \sim e^{S}$ result obtained by Hartle and Hawking was first obtained by Vilenkin, and the $P \sim e^{-S}$ result attributed to Vilenkin was first obtained by Linde.}. 

Herewith, we will consider the above context with the assistance of a new probe, not yet given widespread use. More concretely, a series of  arguments and findings using fractional calculus in quantum physics have been recently presented \cite{par1}. The primary rationale is that if we limit the path integral description of quantum mechanics to Brownian paths exclusively, explaining concrete essential quantum phenomena would be complicated \cite{Laskin1}.
Due to these issues, extended variants of the Schr\"odinger equation (SE), 
have been considered, where derivatives of fractional order, e.g., $\partial^\alpha / \partial x^\alpha$, $1<\alpha\leq 2$, $\alpha$ being a rational number, are used.
Such tool from fractional calculus assists
 fractional quantum mechanics (FQM), where  space-fractional \cite{space}, time-fractional \cite{time}, and space-time-fractional \cite{space-time} variants of the ordinary SE have been the subject of attention.
 Furthermore, in the last few years, FQM has been 
 employed as a means to explore features within 
 quantum field theory and gravity for fractional space-time \cite{FQ1,calcagni2010quantum}, and the fractional quantum field theory at positive temperature \cite{lim2012casimir,lim2006fractional}.
  It has pointed to exciting opportunities; 
  please see  \cite{FQ1,FQ2,FQ3}, for a recent survey. 

The purpose of this paper is therefore to employ a 
particular  fractional WDW equation of the de Sitter (deS) space, embracing it as a probe model. We are aware of the restrictive scope but  absorbing features can be retrieved. More concretely, 
{for a closed de Sitter geometry,} we obtain the fractional tunneling (Linde--Vilenkin) wavefunction and the fractional extension of the no-boundary wavefunction (Hartle--Hawking), followed by computing the corresponding probability distributions. Subsequently comparing those 
expressions with the corresponding usual nucleation rates of the Linde--Vilenkin and 
for the Hartle--Hawking deS wavefunction, our results suggest that a  fractal behavior for the de Sitter horizon can be estimated: the tunneling (Linde--Vilenkin) wavefunction favors fractal dimensions less than $2.5$; whereas in the no-boundary proposal (Hartle--Hawking)
case, the corresponding wavefunction selects a fractal horizon with dimensions close to $3$. Furthermore, after nucleating, the universe may either enter an accelerated power-law phase (Linde--Vilenkin wavefunction) or may instead proceed into a decelerated phase (Hartle--Hawking case). In addition,  our analysis allows to 
discuss  (non-trivial compact) flat and open scenarios. Since the probability of creation of a closed inflationary universe in the tunneling proposal is exponentially suppressed, a flat or an open universe emerges  favored within  fractional inflationary quantum universe.


Let us start by briefly summarizing the canonical quantization of a deS universe.

\section{Canonical quantization of a de Sitter universe}\label{S2}

The minisuperspace ADM action for a deS universe with a positive constant curvature geometry is given by \cite{MonizBook}
\begin{multline}
    \label{2-1}
    S_\text{ADM}=\frac{3\pi}{4 G}\int\Big(-\frac{a(t)}{N(t)}\dot a(t)^2+\\a(t)N(t)-\frac{\Lambda}{3}N(t)a(t)^3\Big)dt,
\end{multline}
where $G$ is the gravitational constant, $\Lambda$ is the cosmological constant, $N(t)$ is the lapse function, and $a(t)$ represents the scale factor. 
The minisuperspace ADM Hamiltonian for the deS universe with a positive constant curvature geometry is then  given by \cite{MonizBook}
\begin{multline}
    \label{2-2}
    H_\text{ADM}=\\-N\left\{\frac{1}{3\pi m_\text{P}^2a(t)}\Pi^2+ \frac{3\pi m_\text{P}^2}{4}a(t)\left(1-\frac{\Lambda}{3}a(t)^2\right)\right\},
\end{multline}
where $m_\text{P}=1/\sqrt{G}$ is the Planck mass \footnote{Throughout  this paper we shall work in natural units, $\hbar=c = k_B = 1$.}, 
and
\begin{equation}
    \label{2-3}
    \Pi=-\frac{3\pi m_\text{P}^2}{2}\frac{a(t)\dot a(t)}{N(t)},
\end{equation}
is the conjugate momentum of the scale factor, $a(t)$.
The WDW equation for the wavefunction of the deS universe is retrieved as  \cite{Darabi}
\begin{equation}
    \label{2-4}
    \left\{-a^{-p}\frac{d}{da}a^p\frac{d}{da} +\left(\frac{3\pi m_\text{P}^2}{2} \right)^2a^2\left(1-\frac{\Lambda}{3}a^2\right)\right\}\Psi(a)=0,
\end{equation}
where the parameter $p$ represents the factor-ordering ambiguity \cite{Vilenkin}. In this paper, we are interested in a 
semi-classical approximation, so we neglect the operator-ordering, and hence we set $p=0$. The WKB Linde--Vilenkin wavefunction of the above one-dimensional WDW equation, disregarding the preexponential factor,
 is \cite{Vilenkin}
\begin{equation}
    \label{2-5}
\psi_\text{LV}(a)=  
\begin{cases}
        e^{\int_a^L|\Pi(a')|da'},~~~~~~~~~~a<L,\\
            e^{- i\int^a_{L}\Pi(a')da'+ i\frac{\pi}{4}},~~~~a\geq L,
\end{cases}
\end{equation}
where $L=\sqrt{3/\Lambda}$ is the deS horizon radius. The wavefunction decays  exponentially toward $a=L$ and after that, it  oscillates,   describing  an expanding deS universe. On the other hand, the Hartle--Hawking wavefunction is \cite{Hawking1}
\begin{equation}\label{Hartle}
  \psi_\text{HH}(x)=   \begin{cases}
            e^{\int_a^L|\Pi(a')|da'},\hspace{2cm}a<L,\\
    \cos\left(\int^a_{L}\Pi(a')da'-\frac{\pi}{4} \right), ~~~~~a\geq L.      
    \end{cases}
\end{equation}
The nucleation rate for the Linde--Vilenkin wavefunction in this  WKB approximation is given by \cite{Vilenkin1,Vilenkin14,Vilenkin15,Vilenkin16}
\begin{multline}
    \label{2-6}
    \mathcal P_\text{LV}    =\frac{|\psi_\text{Tun}(L)|^2}{|\psi_\text{Tun}(0)|^2}\propto e^{-2\int_0^{L}|\Pi(a')|da'}=e^{-\mathcal S_\text{DS}},
\end{multline}
where $\mathcal S_\text{DS}$ is the entropy of deS space-time \cite{Padman}
\begin{equation}
    \label{2-7a}
  \mathcal  S_\text{DS}=\frac{4\pi L^2}{4G}=\frac{A_\text{DS}}{4G},
\end{equation}
in which $A_\text{DS}$ is the area of the horizon. According to the tunneling approach to quantum cosmology, our universe is thought to have begun in a tunneling event. The universe instantly begins a deS inflationary expansion when it nucleates \cite{Vilenkin1}.
On the other hand,  the unnormalized
bare probability density for Hartle--Hawking wavefunction 
can be computed from (\ref{Hartle}), bringing 
\begin{multline}\label{2-7b}
     \mathcal   P_\text{HH}=|\psi(a<L)|^2\propto e^{2\int_0^{L}|\Pi(a')|da'}=e^{\mathcal S_\text{DS}}.
\end{multline}
It can be further added that our deS setting can be taken as a background upon quantum fluctuations may be considered. Albeit 
somewhat descriptive at this 
perspective, 
we can take $\Lambda$ as an effective cosmological constant, associated as  
 $\Lambda_\text{eff}=8\pi GV_0=8\pi V_0/m_\text{P}^2$ to some 
 potential of some field and in a 
 maximum. Quantum fluctuations may 
then push such  field away from the maximum; such  vacuum energy remains essentially constant, $V=V(0):=V_0$, if the potential is suitably flat, and the universe then expands exponentially $a(t)=\exp(t/L)$, where
\begin{equation}\label{2-16}
    L:=\sqrt{\frac{3m_\text{P}^2}{8\pi V_0}}.
\end{equation}
From Eqs.(\ref{2-7a}) and (\ref{2-7b}), it is clear that the tunneling and the no-boundary states lead to different
physical implications. According to the tunneling proposal, the highest probability is obtained for the smallest entropy values. Thus, the tunneling wavefunction `predicts' that the universe is likely to nucleate with the smallest possible entropy. On the contrary,  the Hartle--Hawking probability is peaked at the smallest values of the cosmological constant, and thus the corresponding wavefunction tends to predict initial conditions with maximum possible entropy. 
Or, in other words, the Linde--Vilenkin wavefunction `predicts' that the universe will nucleate with a large vacuum energy \cite{Vilenkin1}. Due to the negative sign in (\ref{2-7b}), the no-boundary state instead increases the contribution of empty universes with $V_0=0$ in the entire quantum state, leading to the seemingly contradictory conclusion that indefinitely huge universes are infinitely more likely than finite-sized universes. The  probability in equation (\ref{2-7a}), on the other hand, promotes large values of $V_0$ capable of causing deS inflationary scenarios. As a result, it appears that the Linde--Vilenkin prescription is physically more alluring than the no-boundary prescription, as far as a deS inflation is remarked in this reduced scope of discussion. 

{Nonetheless, suppose we extend 
{the above as to include an explicit }
inflaton scalar field. In that case, the situation changes drastically. There is a general {domain}
$V_0\leq10^{-8}m_P^4$ derived from the amplitude constraint on gravitational waves generated during inflation in all 
models where the effective potential does not change significantly during the final stages of inflation. Many inflationary models, including new inflation, hybrid inflation
predict {that} inflation will occur at $V_0\ll10^{-8}m_P^4$. 
{Subsequently,}
the minimal size of a closed inflationary universe is given by $1/H=\frac{3m_P}{8\pi\sqrt{V_0}}$. The probability of quantum creation of an inflationary universe is suppressed by a factor of $10^{-10^{10}}$ in a specific example with $V_0=10^{-12}m_P^4$ \cite{Linde:2004nz,Linde:2017pwt}. As 
{remarkably} shown in Refs. \cite{Linde:2004nz,Linde:2017pwt,Coule:1999wg,Linde:2014nna}, {the} tunneling proposal, {allows the creation of a  topologically compact nontrivial} open or flat universe. {Moreover, it seems more 
likely to occur than the emergence of a}  closed universe. Since such processes are unaffected by exponential factors, the universe can be created even if the energy density during inflation is much lower than the Planck density. After a long period of inflation, such universes become indistinguishable from isotropic flat universes.}

\section{de sitter fractional quantum cosmology}

{Recent quantum gravity results have given a significant boost to the growing use of fractional calculus in quantum cosmology. 
In \cite{Lauscher:2005qz, 
Ambjorn:2005db, 
Modesto:2008jz,
Horava:2009if,
Biswas:2011ar}, 
the dimension of space-time changes with scale. Because fractional integro-differential operators can describe these processes, using fractal processes in quantum physics is a precursor to incorporating fractional calculus into quantum theory 
\cite{1966PhRv79N}.
}


Let us now convey our discussion towards a  fractional quantum formulation of deS space-time.
To obtain a fractional WDW equation, let us see how a fractional SE can be expressed. 
{An analogy can be drawn from the diffusion equation, which is extended to the fractional diffusion equation to describe anomalous diffusion in the case of fractional generalization of quantum mechanics. Because the Schr\"odinger equation is identical to the diffusion equation up to a few constants, fractional generalization of the Schr\"odinger equation is also possible. Consider the Hamiltonian of a classical system, $H=\frac{\mathbf{p}^2}{2m}+V(\mathbf{r})$, where $\mathbf{r}$ and $\mathbf{p}$ denote the space coordinates and momentum associated with a particle of mass $m$, respectively. As previously stated, the Brownian-like quantum mechanical trajectories used in Feynman's framework are replaced by L\'evy-like ones in the Laskin approach. Applying a natural generalization of the preceding classical Hamiltonian as $H_\alpha(\mathbf p, \mathbf r) := D_\alpha |\mathbf p|^\alpha+ V(\mathbf r)$, where $1< \alpha\leq 2$ is the L\'evy's fractional parameter and it is associated to the concept of L\'evy path \cite{par1}, $D_\alpha$ is a scale coefficient. Applying the standard canonical quantization procedure, $(\mathbf{r},\mathbf{p})\rightarrow (\mathbf{r},\mathbf{-i\hbar\nabla})$, to this extended Hamiltonian gives the corresponding quantum Hamiltonian of the system $H_\alpha= D_\alpha
(-\hbar^2 \Delta)^{\alpha/2}+V(\mathbf{r})$.  }
{Therefore, to  obtain  a  fractional  extension of the Schr\"odinger equation},
\begin{equation}
 i\hbar \frac{\partial
 \psi(\mathbf{r},t)}{\partial  t}=
-\frac{\hbar^2}{2m}\Delta \psi(\mathbf{r},t)
+ V(\mathbf{r},t) \psi(\mathbf{r},t),
\label{7-0a}
\end{equation}
we may replace the ordinary Laplacian, $\Delta$, with the 
fractional Riesz, $(-\hbar^2 \Delta)^{\alpha/2}$, namely 
\begin{equation}
    -\frac{\hbar^2}{2m}\Delta\longrightarrow D_\alpha
(-\hbar^2 \Delta)^{\alpha/2}.
\end{equation}
The
 fractional Laplacian in the  Riesz form, $(-\hbar^2 \Delta)^{\alpha/2}$,  for 3-dimensional
Euclidean space is defined in terms of the Fourier transformation \cite{Rie}
 \begin{equation}
 \begin{array}{cc}
(-\hbar^2 \Delta)^{\alpha/2} \psi(\mathbf{r},t)
=\mathcal F^{-1}|\mathbf{p}|^\alpha\mathcal F\psi(\mathbf{r},t)\\
\\
=\displaystyle\frac{1}{(2\pi\hbar)^3}
\int d^3 p e^{i\frac{\mathbf{p}\cdot \mathbf{r}}{\hbar}}
|\mathbf{p}|^\alpha
\int e^{-i\frac{\mathbf{p}\cdot \mathbf{r}'}{\hbar}}\psi(\mathbf{r}',t)d^3r',
\end{array}
\end{equation}
in which $\mathbf{p}=\sqrt{p_1^2+p_2^2+p_3^2}$.
The result is the fractional SE
\begin{equation}
    i\hbar \frac{\partial \psi(\mathbf{r},t)}{\partial  t}=
D_\alpha
(-\hbar^2 \Delta)^{\alpha/2} \psi(\mathbf{r},t)
+ V(\mathbf{r},t) \psi(\mathbf{r},t).
\label{7-0}
\end{equation}

Returning to our case study, the supermomentum constraint $H_i$ is identically zero in homogeneous cosmological models, 
and the shift function $N_i$ can be adjusted to zero without neglecting any of Einstein's equations. As a result, the superspace WDW equation is simplified to the minisuperspace WDW equation:
\begin{equation}\label{WDW}
    \left\{\frac{1}{2}\Box+U(q^\nu)\right\}\Psi(q^\nu)=0,~~~\nu=0,...,N-1,
\end{equation}
where $q^\alpha$ are coordinates of $N-$dimensional minisuperspace, $\Box=\frac{1}{\sqrt{-f}}\partial_\alpha(\sqrt{-f}f^{\alpha\beta}\partial_\beta)$ is the  d’Alembertian operator, $f_{\alpha\beta}$ denotes the corresponding minisuperspace metric with signature $(-,+,+,...,+)$ and $U(q^\nu)$ is the  potential.
To obtain a fractional extension of the above WDW equation we may replace the ordinary d'Alembertian operator in (\ref{WDW}) as 
\cite{Rie,Tarasov:2018zjg,Calcagni_2021,Calcagni:2011kn}, namely 
\begin{equation}
(-\Box)^{\alpha/2} \Psi(q^\alpha)
=\mathcal F^{-1}(|\mathbf{p}|^\alpha(\mathcal F\Psi(\mathbf p)), 
\end{equation}
where $|\mathbf p|=\sqrt{-p_0^2+p_ip^i}$, $i=1,2,...,N$, and $\mathcal F$ is a Fourier transformation.
Let us be more clear and specific. The WDW equation (\ref{2-4}) is a one-dimensional ES with zero energy. Thus, to construct a particular fractional WDW equation, we may, similarly to the SE case,  replace the ordinary derivative for the fractional Riesz derivative, namely  \cite{FQ2,FQ3,FracBH}
\begin{equation}
    \label{2-8}
    -\frac{d^2}{da^2}\longrightarrow m_\text{P}^{2-\alpha} (-\frac{d^2}{da^2})^\frac{\alpha}{2},~~~~1<\alpha\leq2.
\end{equation}
We remind that the fractional  Riesz derivative is further defined in terms of the Fourier transformation \cite{Rie}
 \begin{equation}
 \begin{array}{cc}
(-\frac{d^2}{da^2})^{\alpha/2} \Psi(a)
=\displaystyle\frac{1}{2\pi}
\int d\Pi e^{i{\Pi a}}
|\Pi|^\alpha
\int e^{-i{\Pi a'}}\Psi(a')da'.
\label{2ref8}
\end{array}
\end{equation}
Therefore, our fractional WDW equation of a deS space-time is
\begin{equation}
    \label{2-9}
   \left(-\frac{d^2}{da^2}\right)^\frac{\alpha}{2}\Psi(x)+\frac{9\pi^2m_\text{P}^{\alpha+2}}{4}\left(a^2-\frac{\Lambda}{3}a^4\right)\Psi(a)=0.  
\end{equation}
To obtain a WKB approximation, we rewrite the wavefunction  as the exponential of another function, $f$, 
namely 
\begin{equation}
    \label{2-10}
    \Psi(x)=e^{f(x)},~~~~f(x)\in \mathbb C.
\end{equation}
Using (\ref{2-10}) and (\ref{2ref8}) in (\ref{2-9}) and assuming the wavefunction is square-integrable, we find
\begin{equation}
    \label{2-11}
    \frac{d^2f}{dx^2}+\left(\frac{df}{dx} \right)^2+|\Pi|^2=0,
\end{equation}
where $\Pi$ satisfies the following classical fractional superHamiltonian constraint \cite{par1}
\begin{equation}
    \label{2-12}
    |\Pi|^\alpha+\frac{9\pi^2}{4}m_\text{P}^{2+\alpha}a^2\left(1-\frac{a^2}{L^2}\right)=0.
\end{equation}
Within our chosen WKB approximation procedure, we can show  that the corresponding WKB wavefunctions in the 
fractional tunneling or no-boundary case still bear the formal structure as in  (\ref{2-5}) or (\ref{Hartle}).
Of course, the explicit presence of $\alpha$ as induced from the use of the Riesz derivative, conveys 
significant modifications.

 The WKB Linde--Vilenkin wavefunction of the above one-dimensional fractional WDW equation, disregarding the preexponential factor,
 is 
\begin{equation}
    \label{2-5frac}
\psi_\text{LV}(a)\simeq\begin{cases}
        e^{Ca^{D}F(\frac{D}{2},\frac{1-D}{2};1+\frac{D}{2};(\frac{a}{L})^2)},~~~~~~~~~a<L,
        \\
        \\
        e^{-iCa^{D}F(\frac{D}{2},\frac{1-D}{2};1+\frac{D}{2};(\frac{a}{L})^2)+i\frac{\pi}{4}},~~a\geq L,
\end{cases}
\end{equation}
where $C:=(\frac{3\pi}{2})^{D-1}m_\text{P}^D$, and $F(a,b;c;z)$ is the hypergeometric function and $D=1+2/\alpha$. It is likewise possible to write  the Hartle--Hawking wavefunction in terms of hypergeometric functions.

Consider now the  Linde--Vilenkin and Hartle--Hawking probability distributions. 
These are subsequently written as
\begin{equation}
    \label{2-13}
    \mathcal P_\text{LV}=e^{-\left(\frac{3\pi}{4}\right)^{D-1}\frac{\sqrt{\pi}\Gamma(D)}{\Gamma(D+\frac{1}{2})}\left(\frac{3}{8\pi G^2V_0} \right)^\frac{D}{2}},~~
     \mathcal P_\text{HH}={\mathcal P^{-1}_\text{LV}},
\end{equation}
Comparing Eqs.(\ref{2-6}) and (\ref{2-7a}) with the above result, it  suggest us to define a fractional 
entropy \footnote{ In 
Ref.\cite{FracBH}  a  space-fractional formulation  for the Schwarzschild black hole allowed to formulate that the horizon  unveiled a fractal 
structure and 
$\alpha$-dependent. 
The proposed fractional entropy of the black hole is 
$\mathcal S_\text{BH}  = \left(\frac{A_\text{BH}}{4G}\right)^\frac{2+\alpha}{2\alpha} = \frac{A_\text{fractal}}{4G}$,
by means of which we hence defined  $A_\text{fractal}$.
In the herewith deS  space-time, we have proposed a definition of the entropy which is similar: $
    \mathcal S_\text{DS}=\frac{A_\text{DS}}{4G}.
$
We think it is admissible to ask whether the  results for the fractal structure of the event  horizon (and the corresponding entropy) are a general behavior.} 
as
\begin{equation}
    \label{2-14}
 \mathcal   S_\text{Frac}:=\frac{A_\text{Frac}}{4G},
\end{equation}
where 
\begin{multline}
    \label{2-15}
    A_\text{Frac}:=4\sqrt{\pi} \left(\frac{3\pi}{4} \right)^{D-1}\frac{\Gamma(D)}{\Gamma(D+1/2)}l_\text{P}^2\left(\frac{L}{l_\text{P}}\right)^D,
\end{multline}
would be  the fractal area of the deS horizon,  $D$ is the fractal dimension of the horizon 
and $l_\text{P}=1/m_\text{P}$ is the Planck length.
Specifically, in the above expression the  horizon surface is a fractal surface whose dimension, $D$, is dependent in 
 the L\'evy's fractional parameter $\alpha$
\begin{equation}\label{Dimension2}
    D=\frac{2+\alpha}{\alpha},~~~~~2\leq D<3.
\end{equation}
Note that for $\alpha=2$ (or equivalently, $D=2$) the fractal area of the horizon will reduce the original smooth area of the deS space. Eqs.(\ref{2-14}) and (\ref{2-15}) show that increasing the L\'evy's fractional parameter, $\alpha$,  also increases the fractality of the horizon area, which in turns causes the change of the effective
Gibbons--Hawking entropy, making it larger than in the smooth standard case.


{Let us further elaborate on the implications from  (\ref{2-9})-(\ref{2-11}), or instead, just (\ref{2-12}), and 
establish how  the situation in the fractional cases become quite} different. 
Eq.(\ref{2-12}) allows to extract and write that the effective fractional ADM Hamiltonian is given by
\begin{equation}\label{2-17}
    H_\text{ADM}(\alpha)=-N\Big\{\frac{1}{3\pi m_\text{P}^\alpha a(t)}|\Pi|^\alpha+U(a)\Big\},
\end{equation}
where
\begin{equation}
    \label{2-18}
 U(a):=   \frac{3\pi m_\text{P}^2}{4}a(t)\left(1-\frac{8\pi V_0}{3m_\text{P}^2}a(t)^2\right).
\end{equation}
Then, the Hamilton's equations, $\dot a=\partial  H_\text{ADM}(\alpha)/\partial\Pi$, $\dot\Pi=-\partial  H_\text{ADM}(\alpha)/\partial a$ together with the Hamiltonian constraint $\mathcal H(\alpha)= H_\text{ADM}(\alpha)/N=0$ lead us to the fractional Friedmann equations in comoving frame ($N=1$):
\begin{multline}\label{2-19}
 \left(\frac{\dot a}{a}\right)^2=\frac{a^{2(1-D)}}{(D-1)^2}\left(\frac{2\sqrt{G}}{3\pi} \right)^{2(D-2)}\left(\frac{a^2}{L^2}-1 \right)^{3-D},\\       \frac{\ddot a}{a}=\frac{(5-2D)\left(-U\right)^{2-D}}{(D-1)^2(3\pi)^{D-2}a^D}\left( \frac{a^2}{L^2}+\frac{D-2}{5-2D}\right),
\end{multline}
where $L$ is defined by (\ref{2-16}).

\begin{figure}[ht]
\includegraphics[width=6cm]{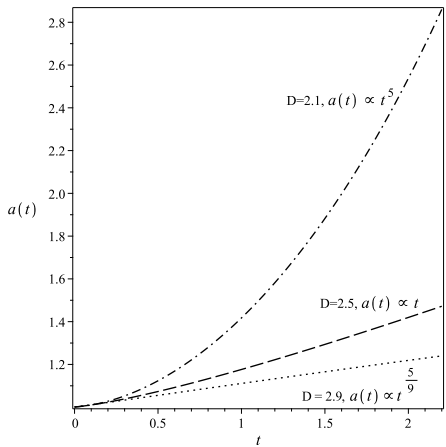}
\caption{Plot of the scale factor of the nucleated  closed universe for three values of fractal dimension,  $D=2.1$ (dash-dot), $D=2.5$ (dash) and $D=2.9$ (dot). The asymptotic behavior of the corresponding scale factors is also indicated in the plot. We used units in which, $m_\text{P}=L=1$. }\label{Fig1}
\centering
\end{figure}

Note as well that for $D=2$ the above field equations reduce the original usual Friedmann equations. Furthermore, as the above field equations show, for $D\neq2$ (or equivalently $\alpha\neq2$, see Fig.(\ref{Fig1})), our  deS model universe does not expand exponentially. The  universe nucleated from nothing and  is accelerated only for $D<2.5$, increasing  as a power-law expansion:
\begin{equation}\label{2-20}
a(t)\propto t^\frac{1}{2(D-2)}.
\end{equation}
The number of e-foldings before the $t_\text{end}$ of a suitable inflation
period and  from a time $t_*$ becomes $N_*=\ln(t_\text{end}/t_*)/(2(D-1))$. For example, for $D=2.001$, we find $N_*=50\ln(t_\text{end}/t_*)>50$;  Nb.  the scale we observe from the cosmic microwave background data conforms to the e-folding in the range $50 < N_* < 60$ \cite{Planck:2013jfk}.
Usually, this kind of power-law expansion may be realized if a scalar field with an exponential potential 
dominates the universe's energy density at an early stage \cite{yokoyama}. As we saw, in the fractional extension of quantum cosmology, for $D\neq 2$, the probability of having a power-law expansion is real and  without such exponential potential. This allows to raise the possibility that 
inflation, an early accelerated stage of expansion, may happen from fractional quantum mechanical perspective. This is of interest and the usual form of the deS inflation will take place only for $D=2$.




Let us add that due to the negative sign in (\ref{2-13}), the no-boundary state  increases not only the contribution of empty universes with $V_0=0$ in the entire quantum state, but also highly fractal horizons with dimension close to three, $D\rightarrow3$. This two effects lead to the seemingly contradictory conclusion that quite huge and decelerating universes are infinitely more likely than finite-sized universes. The tunneling wavefunction probability as  in equation (\ref{2-13}), on the other hand, promotes large values of $V_0$ simultaneously with fractal horizon with dimension close to two, $D\rightarrow2$, and  capable of 
inducing power law accelerated expansion  scenarios. As a result, it appears that the tunneling prescription becomes more alluring within the fractional quantum cosmology herewith, when contrasting with the  no-boundary prescription.

{Similarly to the ending paragraph in the previous section, let us like re-address the points therein introduced. However, we will discuss them extending the context towards fractional quantum cosmology. More precisely, remind}
{
the general {range} $V_0\leq10^{-8}m_P^4$ as derived from the amplitude constraint on gravitational waves generated during inflation. 
From $V_0=10^{-12}m_P^4$, the  probability distribution (\ref{2-13}), {yields}
$10^{-10^{17}}<   \mathcal P_\text{LV}\leq 10^{-10^{10}}$,
where the lower limit is obtained for $D\simeq 3$ and the upper limit in calculated for $D=2$. Thus, the {herewith} closed fractional universe 
{may not be favored}. 
{In fact,} if we consider {instead} creating a compact topologically nontrivial open or flat universes {(see Refs. \cite{Linde:2004nz,Linde:2017pwt,Coule:1999wg,Linde:2014nna}),} then the fractional WDW equation (\ref{2-9}) {becomes}}
\begin{equation}
    \label{jadid1}
           \left(-\frac{d^2}{da^2}\right)^\frac{\alpha}{2}\Psi(x)+\frac{3\mathcal V_k^2m_\text{P}^{\alpha+2}}{16\pi^2}\left(ka^2-\frac{\Lambda}{3}a^4\right)\Psi(a)=0,  
\end{equation}
{where $k=-1,0$, and $\mathcal V_k$ denotes the finite volume of the non-trivial compact spatial hypersurfaces \cite{doi:10.1142/8540}. Besides, the effective fractional ADM Hamiltonian (\ref{2-17}) will generalize to
\begin{multline}\label{jadid2}
    H_\text{ADM}(\alpha)=-N\Big\{\frac{2\pi}{3\mathcal V_km_P^\alpha a}|\Pi|^\alpha+\\\frac{2\mathcal V_km_P^2}{8\pi}a\left(k-\frac{\Lambda}{3}a^2\right)\Big\}.
\end{multline}
The 
fractional Friedmann equation obtained for 
{such universe} 
will be
\begin{equation}\label{jadid3}
     \left(\frac{\dot a}{a}\right)^2=\frac{a^{2(1-D)}}{(D-1)^2}\left(\frac{3\mathcal V_km_P}{4\pi} \right)^{2(D-2)}\left(\frac{a^2}{L^2}-k \right)^{3-D}.
\end{equation}
The semiclassical solution of (\ref{jadid1}) for open universe, $(k=-1)$, disregarding
the pre-exponential factor is
\begin{equation}
    \label{jadid4}
 \Psi(a)\simeq\exp{\pm iCa^{D}F(\frac{D}{2},\frac{1-D}{2};1+\frac{D}{2};-(\frac{a}{L})^2)},
\end{equation}
where $C=(\frac{3\mathcal V_k}{4\pi})^{D-1}m_P^D$, and  a positive sign corresponds to
an expanding universe.
It is easy to show that for flat fractional quantum cosmology, $(k=0)$, the semiclassical wavefunction is given by}
\begin{equation}
    \Psi(a)\simeq a^{-1}\exp{\pm\frac{im_P}{2D-1}\left( \frac{3{\mathcal V}_k^2V_0}{2\pi}\right)^{\frac{D-1}{2}}a^{2D-1}},
\end{equation}
{where a positive sign corresponds to
an expanding universe. Note that for $D=2$ the above expression reduce to the semiclassical wavefunction obtained by Linde in \cite{Linde:2004nz}. This approximation breaks
down at $a\leq (m_P V_0^{(D-1)/2})^{1/(1-2D)}$. Similar to the closed universe, for scale factors much larger than this value, the classical scale factor satisfies (\ref{2-20}). This shows that the {discussion as} proposed in Refs.\cite{Linde:1983mx,Linde:2017pwt}, 
{can also be of value within our appraisal of de Sitter fractional quantum cosmology.}
for the flat universe is also correct for our model. }

\section{Conclusions}

Our simple model brings about a new perspective, whereby  FQM features could have played an influence at the early stages  of  inflation of the universe.  
The properties of an inflaton field are restricted by observations of fluctuations in the CMB and the universe's matter distribution. Despite the fact that the mass of such inflaton and its interaction with matter fields are not determined, well-known arguments favor a heavy (scalar) field with a mass of $10^{13}$ GeV \cite{1983PhLB},  near to the GUT scale, which is frequently used as evidence for the presence of new physics at the junction of the electroweak and Planck scales. {Our findings suggest that  inflation within  fractional quantum cosmology may emerge 
through a different perspective. Namely, taking into consideration the event horizon of the nucleated universe as a fractal surface with dimension $D$. Our analysis is broaden when we consider that  inflation  occurs at energy densities lower than the Planck density. As the  the tunneling proposal indicates an exponentially suppression of the probability of quantum creation of a closed inflationary universe, results suggest that a fractional inflationary quantum universe may 
favour instead a flat or open universes.}

\centerline{\bf Acknowledgments}
This research work was supported by Grants No. Project: UIDB/00212/2020 and 
Project (Programmatic): UIDP/00212/2020 at CMA-UBI (FCT), plus
the COST Action CA18108 (Quantum gravity phenomenology in the multi-me\-ssen\-ger approach).

\bibliography{xsamp}

\end{document}